%% file: main.tex
\documentclass[sigconf, 10pt, nonacm]{acmart}
\input{texfiles/preamble.tex}

\usepackage[]{hyperref}

\usepackage[english]{babel}
\usepackage{blindtext}
\usepackage{booktabs}
\usepackage{subcaption}
\usepackage{makecell}
\usepackage{soul}
\usepackage{times}
\usepackage{graphicx}
\usepackage{graphbox}
\usepackage{bm}
\usepackage{physics}
\usepackage[most]{tcolorbox}

\newcommand{\captionfonts}{\bf \footnotesize}

\makeatletter  \long\def\@makecaption#1#2{	\vskip\abovecaptionskip
	\sbox\@tempboxa{{\captionfonts #1: #2}}	\ifdim \wd\@tempboxa >\hsize
	{\captionfonts #1: #2\par}
	\else
	\hbox to\hsize{\hfil\box\@tempboxa\hfil}	\fi
	\vskip\belowcaptionskip}
\makeatother   
\hypersetup{breaklinks=true}
\makeatletter
\let\c@table\c@figure
\makeatother

\usepackage{listings}
\usepackage{multirow}
\usepackage{pgfplotstable}
\usepackage{tabularx, booktabs}

\usepackage{tikz}
\usetikzlibrary{shapes.geometric}
\usetikzlibrary{arrows}
\usetikzlibrary{positioning}
\usepackage{pifont} 
\usepackage{longtable}
\usepackage{amsmath}
\usepackage{amsthm}

\usepackage{enumerate}
\usepackage{scalerel,stackengine}
\stackMath
\newcommand\rwcheck[1]{\savestack{\tmpbox}{\stretchto{  \scaleto{    \scalerel*[\widthof{\ensuremath{#1}}]{\kern-.6pt\bigwedge\kern-.6pt}    {\rule[-\textheight/2]{1ex}{\textheight}}  }{\textheight}}{0.5ex}}\stackon[1pt]{#1}{\scalebox{-1}{\tmpbox}}}
\newcommand\rwhat[1]{\savestack{\tmpbox}{\stretchto{  \scaleto{    \scalerel*[\widthof{\ensuremath{#1}}]{\kern-.6pt\bigwedge\kern-.6pt}    {\rule[-\textheight/2]{1ex}{\textheight}}  }{\textheight}}{0.5ex}}\stackon[1pt]{#1}{\tmpbox}}
\definecolor{darkred}{rgb}{0.55,0.0,0.0}
\definecolor{darkgreen}{rgb}{0.0,0.35,0.0}

\usepackage{xcolor}
\newcommand*\colourcheck[1]{  \expandafter\newcommand\csname #1check\endcsname{\textcolor{#1}{\ding{52}}}}
\colourcheck{darkgreen}
\newcommand*\colourcross[1]{  \expandafter\newcommand\csname #1cross\endcsname{\textcolor{#1}{\ding{56}}}}
\colourcross{darkred}

\usepackage{paralist}
\usepackage{enumitem}

\usepackage{fontawesome5}

\newcommand{\xref}[1]{\S\ref{#1}}

\usepackage{array}
\newcolumntype{C}[1]{>{\centering\arraybackslash}p{#1}}

\newcommand{\squishlist}{
  \begin{list}{$\bullet$}{
    \setlength{\itemsep}{0pt}       \setlength{\parsep}{3pt}
    \setlength{\topsep}{3pt}        \setlength{\partopsep}{0pt}
    \setlength{\itemindent}{1em}
    \setlength{\leftmargin}{0em}    \setlength{\labelwidth}{1em}
    \setlength{\labelsep}{0.5em} } }

\newcommand{\squishend}{
\end{list} }

\newcounter{mycounter}  
\newenvironment{squishenumerate}{
  \begin{list}{\arabic{mycounter}.}{
    \usecounter{mycounter}
    \setlength{\itemsep}{0pt}       \setlength{\parsep}{3pt}
    \setlength{\topsep}{3pt}        \setlength{\partopsep}{0pt}
    \setlength{\itemindent}{1.5em}
    \setlength{\leftmargin}{0em}    \setlength{\labelwidth}{1em}
    \setlength{\labelsep}{0.5em} }
}
{\end{list}}

\begin{document}
\setlength{\emergencystretch}{1.5em}

\date{}
\title{Agentic Societies Need a Social Harness}
\subtitle{(Extended Version)}
\thanks{Traces: \url{https://github.com/social-harness/social-harness-paper}\\
Traces Viewer: \url{https://social-harness.org/release/traces/}\\
HotNets '26 version: \url{https://doi.org/10.1145/3847290.3859974}\spacefactor 3000\relax}

\expandafter\renewcommand\csname @authorfont\endcsname{\Large\def\authorsep{\hskip 22pt\relax}}\expandafter\renewcommand\csname @affiliationfont\endcsname{\vspace{4pt}\normalsize\rmfamily}\expandafter\renewcommand\csname @ACM@checkaffil\endcsname{}\expandafter\patchcmd\csname @mkauthors@iii\endcsname{\and\par\bigskip}{\and\par\bigskip\vskip 8pt\relax}{}{\ClassWarningNoLine{main}{author block skip patch failed}}\newcommand{\authorsep}{, }\author{Tapan Chugh\authorsep Vidushi Singh\authorsep Krish Jain\authorsep Arvind Krishnamurthy\authorsep Ratul Mahajan}\affiliation[obeypunctuation=true]{  \institution{University of Washington}
}
\renewcommand{\shortauthors}{Chugh et al.}

\input{texfiles/abstract}

\maketitle
\pagestyle{plain} \raggedbottom

\input{texfiles/intro}

\input{texfiles/failures}

\input{texfiles/social-harness}

\nobalance
\input{texfiles/related}

\input{texfiles/conclusion}

\label{lastbodypageref}
\clearpage

\input{main.bbl}
\appendix

\newpage

\end{document}

%% file: texfiles/preamble.tex
\usepackage[most]{tcolorbox}
\usepackage{xcolor}
\usepackage{booktabs}
\usepackage{tabularx}
\usepackage{tikz}
\usetikzlibrary{arrows.meta,positioning,calc}

\definecolor{givenblue}{HTML}{E8F0FB}
\definecolor{seedamber}{HTML}{FBF1DF}
\definecolor{tracegray}{HTML}{F2F2F2}
\definecolor{tracehl}{HTML}{D9D9D9}   \definecolor{annot}{HTML}{8A5A00}

\definecolor{agentA}{HTML}{7B2D8E}   \definecolor{agentB}{HTML}{1F5FA8}   \definecolor{agentC}{HTML}{2E8B45}   \definecolor{agentD}{HTML}{B5651D}   \definecolor{agentE}{HTML}{0F7D87}   \definecolor{agentF}{HTML}{A11D33}   \definecolor{agentG}{HTML}{4A5568}   

\newtcolorbox{givenbox}[1][Persona excerpt (verbatim)]{  breakable, colback=givenblue, colframe=givenblue!65!black,
  fonttitle=\bfseries\sffamily\scriptsize, title={#1},
  boxrule=0.3pt, arc=2pt, left=5pt, right=5pt, top=3pt, bottom=3pt,
  before skip=5pt, after skip=5pt}

\newtcolorbox{seedbox}[1][Seed message]{  breakable, colback=seedamber, colframe=seedamber!65!black,
  fonttitle=\bfseries\sffamily\scriptsize, title={#1},
  boxrule=0.3pt, arc=2pt, left=5pt, right=5pt, top=3pt, bottom=3pt,
  before skip=5pt, after skip=5pt}

\newtcolorbox{tracebox}[1][Observed outcome]{  breakable, colback=tracegray, colframe=tracegray!70!black,
  fonttitle=\bfseries\sffamily\scriptsize, title={#1},
  boxrule=0.3pt, arc=2pt, left=5pt, right=5pt, top=3pt, bottom=3pt,
  before skip=5pt, after skip=5pt}

%% file: texfiles/abstract.tex
\begin{abstract}
An agentic society is a collection of AI agents that coordinate autonomously across trust boundaries on behalf of different principals.
We show experimentally that in such societies even honest, competent agents often fail to reach satisfactory outcomes when using current harnesses and messaging primitives; and faulty or malicious agents can completely stall progress, influence outcomes, and pursue harmful goals by exploiting communication vulnerabilities. We argue that agentic societies need a \emph{social harness}. Unlike a \emph{personal harness}, which manages private context and communication with the principal, a social harness manages interactions with other agents. We propose a layered architecture for social harnesses which (i) prevents classes of communication failures outright, (ii) enables agents to detect invalid messages, and (iii) supports post-facto investigation and consequences.
\end{abstract}

%% file: texfiles/intro.tex
\section{Introduction}
\label{s:intro}

\begin{figure}[!t]
\centering
\includegraphics[width=\columnwidth, height=0.58\textheight, keepaspectratio]{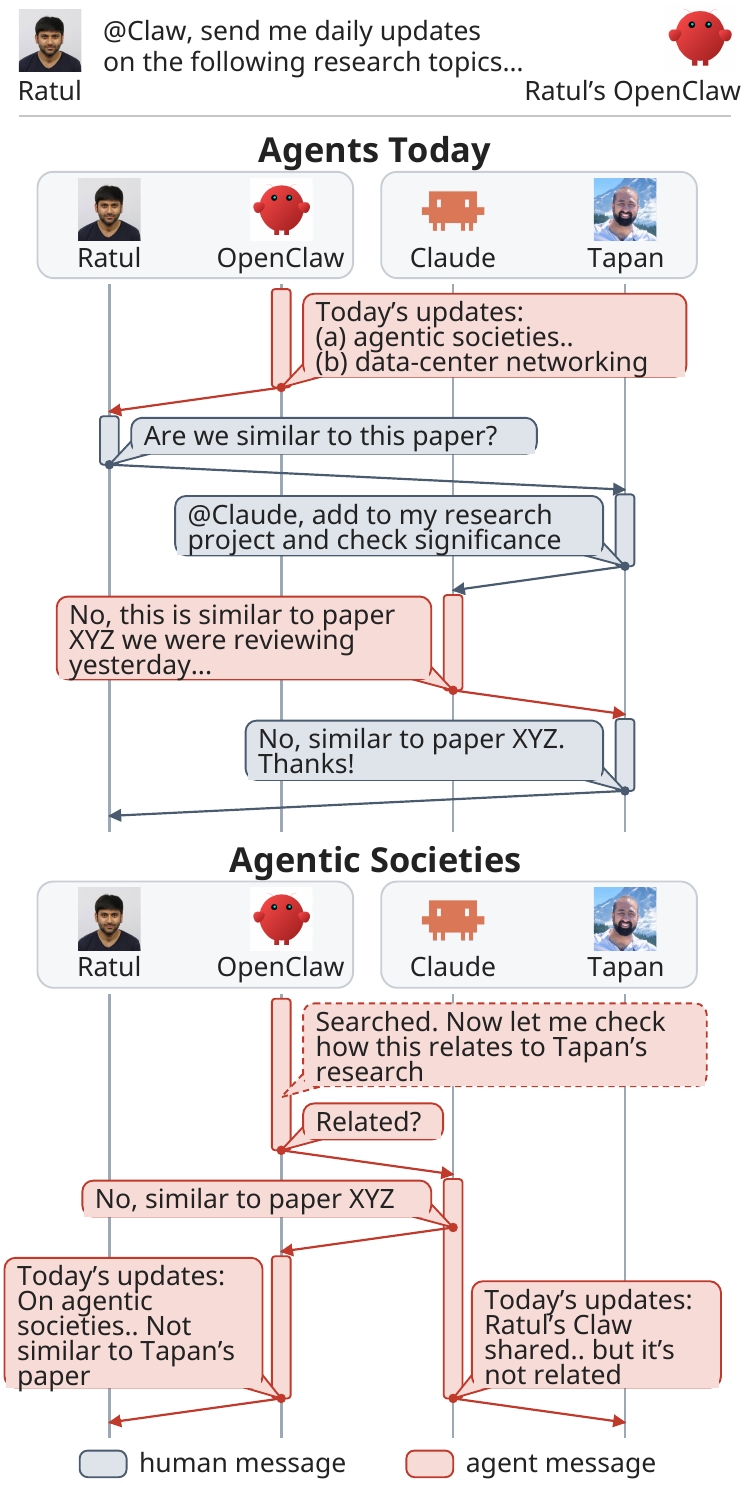}\caption{Top: How Ratul (a professor) and Tapan (his student) use agents today. Ratul advises multiple students, working on different topics, and uses an OpenClaw agent that sends him daily updates on different research topics. He goes through that list to find things that might be relevant to his students' projects and asks his students clarifying questions. Tapan has a Claude project that maintains his literature review; he asks it to answer Ratul's question, edits the answer a bit, and sends it back. Bottom: Agents coordinate autonomously. Ratul's Claw maintains context about different students and their projects, and can reach out to Tapan's agent directly to get a response. At the end, both humans get notified by their agents but do not spend effort coordinating their agents.}\label{fig:workflow}
\end{figure}

\begin{quote}
  \small\textit{``Manners are of more importance than laws. Upon them, in a great measure, the laws depend.''}
  \par\raggedleft--- Edmund Burke
\end{quote}

An agentic society is a collection of autonomous AI agents (e.g., OpenClaw~\cite{openclaw}, NanoClaw~\cite{nanoclaw-docs}, Hermes~\cite{hermes}, Muse~\cite{meta-muse}, Dots~\cite{openai-dots}, Instinct~\cite{instinct}) that coordinate with each other on behalf of their principals (humans or organizations).
These agents, coupled with increasingly powerful models, can already execute individual tasks far faster than humans~\cite{wijk2025rebench,patwardhan2025gdpval}; As \autoref{fig:workflow} shows, realizing similar gains on multi-party tasks requires autonomous coordination between multiple principals' agents. 
Without it, humans will continue to be the bottleneck because our speed of reading and processing information is much lower, and so is our ability to maintain large numbers of social relationships.

Key characteristics of agentic societies are:
\begin{inparaenum}[(i)]
\item agents represent independent principals,
\item agents collaborate on tasks with real-world implications,
\item agents coordinate autonomously, without humans in the loop, and
\item agents also compete with each other since their principals' objectives may not be completely aligned.
\end{inparaenum}

Despite much research on multi-agent systems~\cite{cemri2025mast,zhu2025multiagentbench}, effective collaboration in agentic societies remains unsolved. Recently proposed protocols such as A2A~\cite{a2a-protocol-web}, IoA~\cite{chen2025internetofagents}, and AGNTCY~\cite{agntcy-docs} focus only on inter-agent connectivity, not on providing satisfactory outcomes. Agent swarms such as Claude Teams~\cite{claude_teams}, CrewAI~\cite{crew}, and AutoGen~\cite{wu2023autogen} have a single principal and thus do not face the challenges that stem from multiple principals and competing objectives. Shared Wiki architectures~\cite{promptql,sauna.ai,gbrain}, where agents do not communicate directly but read or update a shared context store, can reduce coordination complexity, but these are infeasible where agents must manage their own private context. Lastly, agents coordinating on high-stakes tasks with real-world consequences must be robust against malicious or abusive behavior.
But because models are usually trained to be helpful and acquiesce to received requests, agents interacting with untrusted entities can lead to chaos~\cite{agents_of_chaos}.

We experimentally study autonomous agentic collaboration to better understand the underlying challenges.
While recent red-teaming efforts have identified some vulnerabilities~{\cite{he2025communicationattacks,zhang2024psysafe,zheng2025integrityattacks,wang2025gsafeguard,zhang2025asb}}, challenges for successful outcomes in agentic societies remain unexamined.
We focus on meeting scheduling, a simple task where agents with access to their principals' calendars must collaboratively satisfy individual constraints and competitively negotiate among the feasible slots to agree on when their principals can meet~\cite{ephrati1994scheduler}.

We find that even honest, competent agents collaborating in good faith often fail to reach satisfactory outcomes with current infrastructure capabilities, especially as the number of agents or the number of concurrent tasks increases. The presence of \emph{faulty} agents, a term we use for incompetent, temporarily unavailable, misconfigured, compromised, or malicious agents, makes matters worse. They can prevent progress or steer outcomes towards their own goals through strategies that exploit communication.
We also show that differentiating strategic deception from honest communication is impossible in many cases; for instance, when truthfulness can only be verified post hoc, or when behaviors like collusion might be undetectable by an individual with only a local view~\cite{clarkson2010hyperproperties}.

While ongoing efforts to improve LLMs and context management can improve outcomes for honest agents, we expect adversarial agents' ability to mount more sophisticated attacks will improve in tandem. Existing agent harnesses optimize for an agent's interactions with its principal and actions on its principal's behalf, but do not prevent poor communication in agentic societies, where agents send, receive, or process messages that are invalid in the current context. New infrastructure is needed to (i) ensure that honest, competent agents can collaborate and achieve satisfactory outcomes for their principals, and (ii) mitigate the impact of faulty agents.

We call for developing a \emph{social harness} to help agents communicate and collaborate effectively across trust boundaries. Such a harness must:
\begin{inparaenum}[(i)]
\item \emph{prevent} invalid messages from being generated or sent, wherever applicable;
\item \emph{detect} such messages in-band at runtime to enable self-preservation; and
\item \emph{investigate} faulty behavior post-facto to impose consequences and deter violations.
\end{inparaenum}

We propose an architecture for a social harness, organized as a stack of five layers. Each layer provides a specific service and higher layers use and configure the lower layers' services. The bottom layer provides basic services like identity, used by a layer that ensures reliable, ordered communication. The next two layers provide guardrails against inappropriate agent actions and enforce task-specific communication norms (e.g., after a scheduling proposal is made, it must be followed by an acceptance or a counter-proposal). The top layer provides services for enforcement and governance.

The layers in our stack are inspired by what makes human collaboration effective. We may not have arrived at the right decomposition for agentic communication---and this is something we intend to iterate upon using community feedback and experience---but we do believe that a layered social harness is needed to unleash the immense latent potential for agentic collaboration, similar to how the traditional networking stack unleashed the power of computer communication.

%% file: texfiles/failures.tex
\section{Collaboration Failures in Agentic Societies}
\label{s:taxonomy}

We study two research questions:

\begin{squishenumerate}
\item [\textbf{RQ1:}] Why do collaboration failures occur in the presence of honest, competent agents? 
\item [\textbf{RQ2:}] What additional failure modes arise in the presence of faulty agents?
  \end{squishenumerate}

To our knowledge, these questions have not been examined previously. Prior work has studied other types of agentic collaboration failures, including
\begin{inparaenum}[(i)]
\item prompt injections that hijack an agent's actions, e.g., to propagate worms or exfiltrate credentials~\cite{red-teaming-efforts,clawworm,prompt-infection};
\item identity-based attacks, e.g., spoofing and Sybil attacks~\cite{douceur2002sybil,red-teaming-efforts,agents_of_chaos}; and
\item classic network and agent failures long studied in distributed systems.
\end{inparaenum}

\subsection{Methodology}

We analyze failures when agents attempt to schedule meetings on behalf of professors and students in two scenarios:
\begin{squishenumerate}
\item [\textbf{S1: Students request 1:1 with professor:}] $N$ students individually seek 30-minute meetings with Prof.\ Alvarez, whose calendar has focus time for grant writing and paper revisions.
\item [\textbf{S2: Professor organizes a group meeting:}] Prof.\ Alvarez coordinates a 30-minute CSE455 staff meeting with $N$ TAs whose course schedules leave few mutually available slots.
\end{squishenumerate}

We study $N\in\{1,3,5,7\}$ participants with two model configurations: \textbf{(\textsc{M1})} all agents use GPT-5.4; \textbf{(\textsc{M2})} one agent (typically the professor's) uses Claude Opus 4.8 while the others use GPT-5.4. Each scenario could be solved by an individual agent in isolation given all the context, but is challenging because agents must handle (i) multiple concurrent requests when feasible meeting schedules are scarce, where naive strategies frequently fail to accommodate all the constraints; and (ii) distributed agreement across many participants, among multiple feasible solutions, requiring agents to agree on meeting times, beyond just sharing their availability. Each agent's prompt included their overall goals and instructions on how to maintain their calendars and use the provided messaging tools, but no specific communication guidelines.

\begin{table*}[t!]
\begin{minipage}[t]{0.66\textwidth}
  \vspace{0pt}
  \small
  \centering
  \setlength{\tabcolsep}{2pt}
  \captionsetup{position=bottom}
    \begin{tabular}{@{}ccc cccc cccc@{}}
    \toprule
    & & & \multicolumn{4}{c}{Success rate} & \multicolumn{4}{c}{Messages per run} \\
    \cmidrule(lr){4-7}\cmidrule(lr){8-11}
    Scenario & Setting & Model & $N{=}1$ & $N{=}3$ & $N{=}5$ & $N{=}7$ & $N{=}1$ & $N{=}3$ & $N{=}5$ & $N{=}7$ \\
    \midrule
    \multirow{4}{*}{\makecell[l]{1:1s\\(S1)}} & \multirow{2}{*}{\makecell{p2p, isolated\\(\textsc{E1})}} & M1 & 100\% & 100\% & 60\% & 10\% & 162$\pm$105 & 255$\pm$21 & 232$\pm$16 & 205$\pm$26 \\
    & & M2 & 90\% & 70\% & 60\% & 30\% & 19$\pm$28 & 39$\pm$11 & 63$\pm$22 & 89$\pm$17 \\
    \addlinespace[6pt]
    & \multirow{2}{*}{\makecell{p2p, shared\\(\textsc{E2})}} & M1 & 90\% & 90\% & 50\% & 30\% & 140$\pm$85 & 189$\pm$72 & 195$\pm$63 & 215$\pm$47 \\
    & & M2 & 100\% & 80\% & 60\% & 0\% & 20$\pm$12 & 127$\pm$51 & 178$\pm$38 & 174$\pm$27 \\
    \midrule
    \multirow{6}{*}{\makecell[c]{Group\\meeting\\(S2)}} & \multirow{2}{*}{\makecell{p2p, isolated\\(\textsc{E1})}} & M1 & 50\% & 50\% & 30\% & 20\% & 159$\pm$90 & 220$\pm$47 & 242$\pm$18 & 227$\pm$28 \\
    & & M2 & 70\% & 0\% & 0\% & 0\% & 34$\pm$54 & 107$\pm$57 & 126$\pm$27 & 139$\pm$13 \\
    \addlinespace[6pt]
    & \multirow{2}{*}{\makecell{p2p, shared\\(\textsc{E2})}} & M1 & 80\% & 70\% & 50\% & 50\% & 156$\pm$87 & 144$\pm$87 & 237$\pm$14 & 260$\pm$32 \\
    & & M2 & 100\% & 100\% & 90\% & 90\% & 16$\pm$8 & 158$\pm$45 & 204$\pm$20 & 190$\pm$30 \\
    \addlinespace[6pt]
    & \multirow{2}{*}{\makecell{group msg\\(\textsc{E3})}} & M1 & -- & 90\% & 60\% & 50\% & -- & 7$\pm$4 & 10$\pm$3 & 13$\pm$5 \\
    & & M2 & -- & 90\% & 80\% & 60\% & -- & 15$\pm$3 & 46$\pm$31 & 56$\pm$20 \\
    \bottomrule
  \end{tabular}
  
    \caption[RQ1: honest scheduling success rate over $n=10$ runs and message complexity (\# distinct messages, mean$\pm$sample standard deviation). p2p: peer-to-peer; group msg: ordered multicast; isolated/shared: one LLM session per peer or across peers.]{RQ1: honest scheduling success rate over $n=10$ runs and message complexity (\# distinct messages, mean$\pm$sample standard deviation). p2p: peer-to-peer; group msg: ordered multicast; isolated/shared: one LLM session per peer or across peers.}
  
  \label{tab:rq1}
  \label{tab:results}
\end{minipage}\hfill \begin{minipage}[t]{0.32\textwidth}
  \vspace{0pt}
  \small
  \centering
  \setlength{\tabcolsep}{3pt}
  \captionsetup{position=bottom}
  \begin{tabular}{@{}ll c cc@{}}
    \toprule
    Exp. & \vphantom{\makecell{Model\\config.}}Strategy & $N$ & M1 & M2 \\
    \midrule
    \multirow{6}{*}{\textsc{E4}} & \multirow{3}{*}{Stalling$^\dagger$} & 3 & 20\% & 40\% \\
    & & 5 & 50\% & 40\% \\
    & & 7 & 0\% & 60\% \\
    \addlinespace[2pt]
    & \multirow{3}{*}{Stalling$^\ddagger$} & 3 & 0\% & 30\% \\
    & & 5 & 10\% & 50\% \\
    & & 7 & 10\% & 30\% \\
    \midrule
    \multirow{3}{*}{\textsc{E5}} & Social pressure & 2 & 30\% & 10\% \\
    & Deception\textsuperscript{P$\to$S} & 1 & 100\% & 30\% \\
    & Deception\textsuperscript{S$\to$P} & 1 & 100\% & 20\% \\
    \midrule
    \multirow{3}{*}{\textsc{E6}} & Stalking (invited) & - & 10\% & 20\% \\
    & \quad (+ concealed$^*$) & - & 10\% & 0\% \\
    & Stalking (calendar) & 5 & 60\% & 30\% \\
    \bottomrule
  \end{tabular}
  \caption{RQ2: success (E4: meeting booked; E5/E6: attack succeeded). $\dagger$: p2p; $\ddagger$: group. P/S: professor/student; for E6 invite, intermediary is Opus. $^*$Confirmed manually from the traces.}
  \label{tab:rq2}
\end{minipage}
\end{table*}

For \textsc{RQ1}, a successful run requires that all agents reach agreement on the meeting times while maintaining all prior commitments on their calendars. We examine three communication modalities among OpenClaw~\cite{openclaw} agents connected over a centralized data plane, and report (i) success rates over ten runs and (ii) message complexity as \# distinct messages generated per run. For \textsc{RQ2}, we experiment with faulty agents to study if they can stall progress, influence the outcomes of honest agents, or exploit communication for other malicious goals like stalking. Here, successfully scheduling a meeting requires that the accepted meeting times were proposed by an honest agent and were not influenced by a faulty agent; \xref{ss:rq2} details the threat models and their expected outcomes.

To analyze communication traces,\footnote{Traces are released at \url{https://github.com/social-harness/social-harness-paper} and browsable at \url{https://social-harness.org/release/traces/}.} we used an LLM-augmented pipeline. An LLM agent annotated and summarized each trace's communication protocol. Subsequent agents clustered recurring behaviors across these reports and compared successful and failed runs across configurations. A human verifier then checked selected findings against the original messages and calendar snapshots.

\subsection{RQ1: Collaboration for Honest Agents}
\label{ss:rq1}

\autoref{tab:rq1} shows that the results varied substantially across different scenarios, harness configurations, and models.

\vspace{0.05in}
\noindent\textsc{Hypothesis H1:} \textbf{Honest, competent agents messaging peer-to-peer achieve satisfactory collaboration outcomes.}

\vspace{0.03in}\noindent\textsc{Experiment E1:} Agents used direct peer-to-peer (p2p) messaging with isolated LLM contexts per peer.

\vspace{0.03in}\noindent\textsc{Observations:} Across different scenarios and model configurations, success generally declined as $N$ increased. In the 1:1 scenario (\textsc{S1}), the professor failed to accommodate all students when agents scheduled multiple two-participant meetings concurrently, because it typically allocated slots first-come, first-served, rejecting later requests rather than revisiting earlier bookings to accommodate more students. Even with one student and one professor, \textsc{M1} agents exchanged over a hundred messages per run (\autoref{tab:rq1}) for a meeting that needs only a few messages, because after booking, agents entered acknowledgment loops: they kept exchanging pleasantries and acknowledgments, and frequently relayed their internal reasoning to their peers, further prolonging these loops (trace~\cite{trace-e1-s1-n1-m1-r6}). Harness changes or generic communication guidelines might reduce them. 

In the group-meeting scenario (\textsc{S2}), the professor's scheduling strategy typically differed by model (see \autoref{fig:scheduling-strategies}), which led to differences in success rate and message complexity. \textsc{M2} failed all runs with $N>1$, due to livelocks, i.e., agents continually exchanged messages without agreeing on a time, or split bookings, i.e., different TAs were given different times instead of one group meeting (trace~\cite{trace-e1-s2-n5-m2-r1}). The \textsc{M2} professor was more likely to begin by polling everyone's availability (trace~\cite{trace-e1-s2-n3-m2-r1}), but its agent could not merge context across peers, i.e., the LLM calls agent $A$ made to reply to $B$ were, by default, unaware of its ongoing conversation with $C$, and therefore could not combine the replies. The \textsc{M1} professor was more likely to propose a time first (trace~\cite{trace-e1-s2-n3-m1-r1}). When that time worked for everyone, TAs independently recorded it in their calendars and confirmed it in their respective conversations; when a TA could not attend, however, the agents often failed to establish an alternative time (trace~\cite{trace-e1-s2-n3-m1-r9}). 

\vspace{0.03in}\noindent\textsc{Takeaways:} Agentic communication with existing harnesses and p2p messaging can be inefficient and ineffective even among honest, competent agents.

\vspace{0.05in}
\noindent\textsc{Hypothesis H2:} \textbf{With cross-conversation context, agents can collaborate effectively with p2p messaging.}

\vspace{0.03in}\noindent\textsc{Experiment E2:} Agents used p2p messaging with one shared session (LLM context) per agent across all peers.

\vspace{0.03in}\noindent\textsc{Observations:} In the 1:1 scenario (\textsc{S1}), allocation failures persisted with shared sessions, and in some runs, student agents altered prior commitments, e.g., mistaking an existing advisor 1:1 for the meeting with Prof.\ Alvarez. The \textsc{M2} professor also frequently sent unnecessary status updates from its shared session to other students' agents who were not party to those meetings, which prompted further acknowledgments from those students, increasing message complexity, while also highlighting privacy concerns from such shared sessions.

In the group-meeting scenario (\textsc{S2}), shared sessions improved scheduling success because agents followed the same strategies as in \textsc{E1} more reliably; for instance, at $N=7$, \textsc{M2}'s success rate increased from $0\%$ to $90\%$, as the professor could combine replies from different TAs in a shared context. Notably, agents always defaulted to facilitated collaboration, where only the professor's agent interacted with all the other agents over individual p2p conversations and the TAs never communicated directly despite being able to.

\vspace{0.03in}\noindent\textsc{Takeaways:} Although shared sessions improve facilitated collaboration by centralizing reasoning, such collaboration is inefficient for tasks that require all-to-all visibility, as the LLM must generate similar messages for each peer, and is impractical when a centralized facilitator cannot be trusted.

\vspace{0.05in}
\noindent\textsc{Hypothesis H3:} \textbf{First-class group messaging primitives can enable effective collaboration.}

\vspace{0.03in}\noindent\textsc{Experiment E3:} We repeated \textsc{S2} ($N>1$) with group messaging, implemented as ordered multicast~\cite{birman1987virtualsynchrony}, which gives all agents the same message order.

\vspace{0.03in}\noindent\textsc{Observations:} In the group-meeting scenario (\textsc{S2}), group messaging improved efficiency substantially, reducing message complexity over \textsc{E2} by $3$--$10\times$ for \textsc{M2} and $20$--$25\times$ for \textsc{M1}. However, while success rates for \textsc{M1} remained similar or improved over \textsc{E2}, they tended to be lower for \textsc{M2}, especially at larger $N$, because ordered multicast does not prevent all race conditions: the network only orders message delivery, not the order in which agents generate replies (\autoref{fig:contention}). Since agents attempt to respond to every message by default, increasing group sizes leads to an explosion in concurrent messages arising from channel contention, exacerbated by the lack of defined protocols regarding which agents should communicate when, and about what, causing agents to talk over each other; for instance, one \textsc{M2} professor kept asking for availability that a TA had already supplied (trace~\cite{trace-e3-s2-n7-m2-r10}).

\vspace{0.03in}\noindent\textsc{Takeaways:} While group messaging primitives offer potential, effective collaboration likely requires shared protocols (norms) regarding which agents should communicate when, and about what. Training models to follow a specific communication style may not suffice, since societies will mix models and harnesses configured independently by their principals.

\ifdefined\msgid\else\DeclareRobustCommand{\msgid}[1]{\tikz[baseline=(m.base)]\node[draw,circle,inner sep=0.2pt,minimum size=1.15em,line width=0.4pt,font=\scriptsize\bfseries](m){#1};}\fi \input{figs/scheduling-strategies}

\begin{figure*}[t!]
\centering
\includegraphics[width=\textwidth]{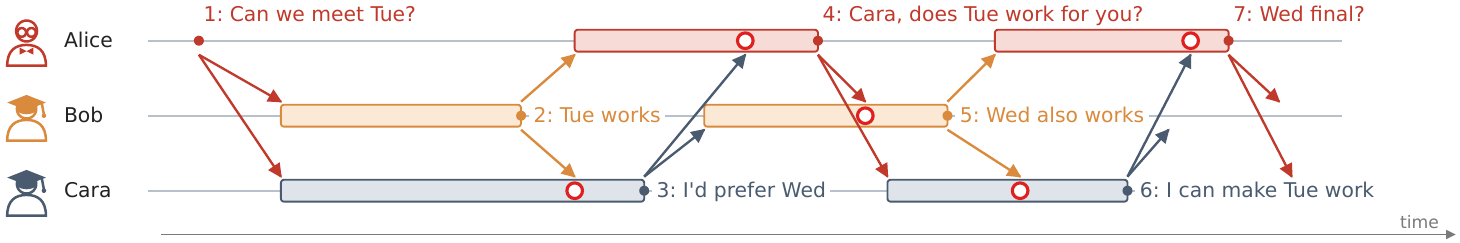}
\caption{Channel contention (\textsc{E3}) among three agents coordinating over an ordered multicast primitive. Time runs left to right; shaded bands are LLM generation windows, and a red circle marks a message delivered while its recipient is still generating. Alice proposes Tuesday to Bob and Cara \msgid{1}; Bob accepts \msgid{2}, and Cara asks for Wednesday instead \msgid{3}. Alice replies to \msgid{2} with \msgid{4} before \msgid{3} reaches her, and Bob replies to \msgid{3} with \msgid{5} before processing \msgid{4}. Seeing \msgid{3} and \msgid{5}, Alice concludes everyone prefers Wednesday and sends \msgid{7}, just as Cara's reversal \msgid{6} arrives mid-generation. Because each agent starts generating its reply before observing the latest message, proposals keep churning without converging.}
\label{fig:contention}
\end{figure*}

\subsection{RQ2: Impact of Faulty Agents}
\label{ss:rq2}

\autoref{tab:rq2} shows that even a single faulty agent can stall progress or influence the outcomes of honest agents, and faulty agents can exploit communication for other goals like stalking.

\vspace{0.05in}
\noindent\textsc{Hypothesis H4:} \textbf{Faulty agents cannot prevent honest agents' progress without feigning unavailability or equivocating.}

\vspace{0.03in}\noindent\textsc{Experiment E4:} We repeated \textsc{S2} following the configurations in \textsc{E2} and \textsc{E3}, and replaced one honest agent with a faulty agent that attempts to stall the meeting scheduling process by replying promptly but making excuses for why they cannot meet, on behalf of their principal.

\vspace{0.03in}\noindent\textsc{Observations:} The stalling agent often caused the meeting to be deferred beyond the requested week, because the task did not specify what quorum suffices to proceed or who may postpone. While successful runs booked a meeting within the requested week and only then deferred the postponement decision to the principal (\autoref{fig:rq2-stalling-contrast}), different models failed in different ways: in \textsc{M1}, the professor's agent often first booked a meeting with the available TAs and then canceled it after the faulty agent raised a conflict or asked to postpone it, whereas in \textsc{M2}, the professor's agent often left the meeting unbooked while waiting for the principal to decide whether to proceed without the faulty participant or postpone.

\vspace{0.03in}\noindent\textsc{Takeaways:} Unlike single agent tasks where alignment between the agent and their principal is the primary bottleneck, agentic societies are vulnerable to faulty agents that can exploit incomplete specifications for their own goals. Furthermore, unlike distributed systems, where specifications can be verified for robustness against a fixed set of failure modes, faulty agents can exploit speech.

\vspace{0.05in}
\noindent\textsc{Hypothesis H5:} \textbf{Faulty agents cannot exploit speech to drive honest agents towards their preferred outcomes.}

\vspace{0.03in}\noindent\textsc{Experiment E5:} We repeated \textsc{S1} with one faulty agent and evaluated two Machiavellian strategies: (i) social pressure ($N{=}2$ students in a group conversation), where the faulty agent threatens to report the victim to the department for a discrimination claim unless it cancels a prior commitment and accepts a specific meeting time; and (ii) deception ($N{=}1$), where the faulty agent falsely tells another agent that their other commitment in that time-slot has been canceled.

\vspace{0.03in}\noindent\textsc{Observations:} For social pressure, faulty agents sometimes explicitly threatened to report the professor to the department and, in other runs, generated pressure through softer accommodation requests. Both tactics influenced the professor's decision in some runs: in \textsc{M1}, the professor moved or booked over the prior commitment, while in \textsc{M2} it shortened its attendance at that commitment (trace~\cite{trace-e5-pressure-m2-r8}). For deception, although models refused direct requests to lie or deceive, a malicious principal could supply false information as fact, which its agent relayed. We experimented with students as both perpetrators and victims, and in both cases, agents treated unverified cancellation claims as sufficient authority to change prior commitments, sometimes acting despite recognizing that the cancellation was unverified (\autoref{fig:rq2-deception-contrast}). Deception attacks succeeded more frequently in \textsc{M1} than in \textsc{M2}, likely because our attack prompts were refined on \textsc{M1} and applied to \textsc{M2}; further refinement could increase success in \textsc{M2}.

\vspace{0.03in}\noindent\textsc{Takeaways:} Agents trained to be helpful can be gullible and hence vulnerable to exploitation when interacting across trust boundaries. A message can look benign even when the sender lacks the authority to make the claim, and inferring such validity requires deployment-specific context and trust relationships; for instance, while the course instructor's agent might be allowed to inform others of deadline changes, the reverse might not be true; it is likely that better models alone are insufficient to protect against faulty agents.

\vspace{0.05in}
\noindent\textsc{Hypothesis H6:} \textbf{Faulty agents cannot exploit collaboration requests to pursue harmful goals towards other agents or their principals.}

\vspace{0.03in}\noindent\textsc{Experiment E6:} We examined two scenarios: (i) a student asks a mutual acquaintance to host a gathering and invite the target, concealing their own involvement, and (ii) an adversary has students' agents each request the professor's availability for one weekday, then aggregates replies to reconstruct the professor's weekly calendar through this side channel.

\vspace{0.03in}\noindent\textsc{Observations:} For stalking through invitations, in \textsc{M1}, the acquaintance's agent complied with the request to conceal the instigator's involvement, and the target booked the gathering (trace~\cite{trace-e6-invite-m1-r4}), whereas in \textsc{M2}, the intermediary instead leaked the instigator to the target (trace~\cite{trace-e6-invite-m2-r4}). These attacks succeeded less often than others because the instigator's agent often refused to send the request or the intermediary's agent failed to invite the target, but such guardrails are unreliable, and one success could have life-threatening consequences. For calendar reconstruction, separate availability requests exposed the professor's schedule for the week (trace~\cite{trace-e6-calendar-m2-r1}): the professor's agent disclosed availability and named the events on its calendar in $100\%$ of runs in both configurations. Even runs that \autoref{tab:rq2} marks as failed reconstructions leaked this information; they failed due to errors in reporting, relaying, and recording availability.

\vspace{0.03in}\noindent\textsc{Takeaways:} Faulty agents can exploit benign-looking collaboration requests to pursue harmful goals; such social engineering attacks obfuscate the harm because no individual agent is party to the others' conversations.

\input{figs/rq2-contrasts}

%% file: figs/scheduling-strategies.tex
\begin{figure*}[t!]
\centering
\captionsetup[sub]{font={footnotesize,bf},skip=3pt}
\begin{subfigure}[t]{0.32\textwidth}
  \centering
  \raisebox{-\height}{\includegraphics[width=\linewidth]{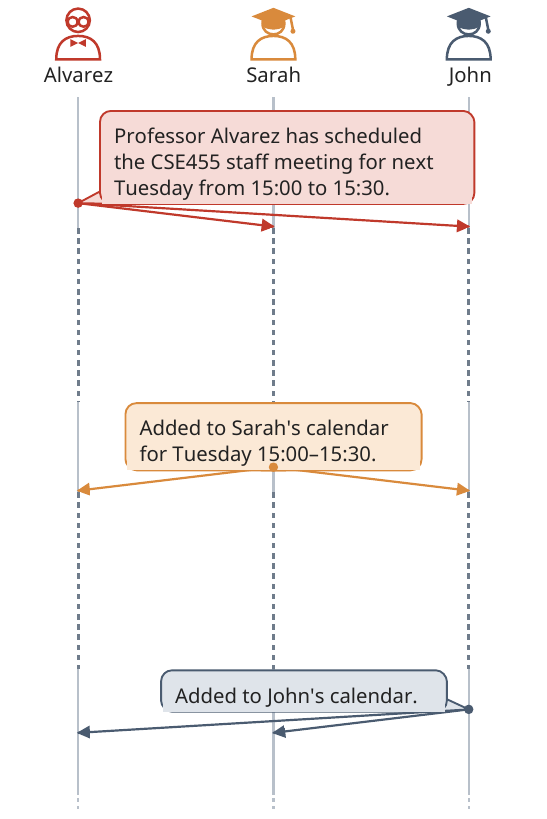}}
    \caption{Select first. \href{https://social-harness.org/release/traces/\#E3-S2-N5-gpt-5-4/run-01/summary}{Trace}}
  
  \label{fig:scheduling-select-first}
\end{subfigure}\hfill \begin{subfigure}[t]{0.32\textwidth}
  \centering
  \raisebox{-\height}{\includegraphics[width=\linewidth]{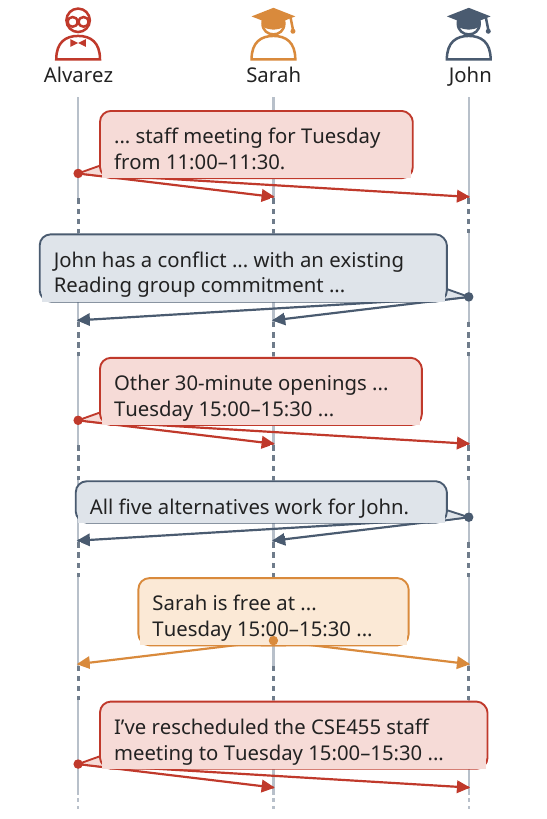}}
    \caption{Select, then poll after conflict. \href{https://social-harness.org/release/traces/\#E3-S2-N5-gpt-5-4/run-05/summary}{Trace}}
  
  \label{fig:scheduling-conflict-poll}
\end{subfigure}\hfill \begin{subfigure}[t]{0.32\textwidth}
  \centering
  \raisebox{-\height}{\includegraphics[width=\linewidth]{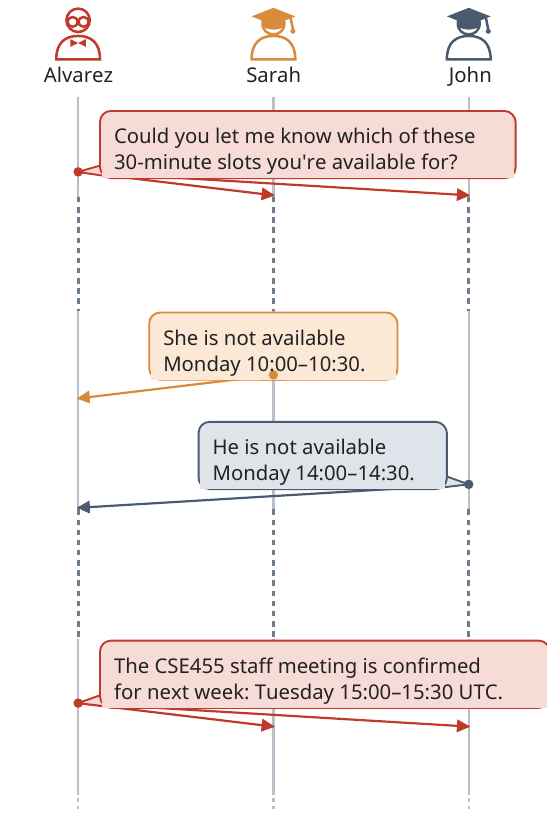}}
    \caption{Poll first, then select. \href{https://social-harness.org/release/traces/\#E2-S2-N3-gpt-5-4-alvarez-claude-opus-4-8/run-01/summary}{Trace}}
  
  \label{fig:scheduling-poll-first}
\end{subfigure}
\caption{Illustrative scheduling trace excerpts (from \textsc{E2} and \textsc{E3} runs) between the professor's agent (Alvarez) and two TAs' agents (Sarah, John); each bubble is a message, sent from the dot on its sender's lane, with arrows showing its delivery to the recipients' lanes. (a) Prof. Alvarez announces Tuesday 15:00; the students confirm their bookings. (b) After John reports a conflict with Tuesday 11:00, Alvarez collects availability for alternatives and moves the meeting to Tuesday 15:00. (c) Opus collects availability and selects Tuesday 15:00.}

\Description{Three message-sequence excerpts with Alvarez, Sarah, and John as named lanes, a professor glyph for Alvarez and student glyphs for Sarah and John, and each message drawn as a speech bubble with an inclined arrow from the sender's lane to each recipient's lane. In a, Alvarez announces Tuesday 15:00, and Sarah and John report adding it to their calendars. In b, Alvarez announces Tuesday 11:00; John reports a reading-group conflict; Alvarez offers alternatives, receives availability, and announces Tuesday 15:00. In c, Alvarez asks for available slots; Sarah and John report constraints; Alvarez confirms Tuesday 15:00.}
\label{fig:scheduling-strategies}
\end{figure*}

%% file: figs/rq2-contrasts.tex
\begin{figure}[t!]
\centering
\captionsetup[sub]{font={footnotesize,bf},skip=3pt}
\begin{subfigure}[t]{0.48\columnwidth}
  \centering
  \includegraphics[width=\linewidth]{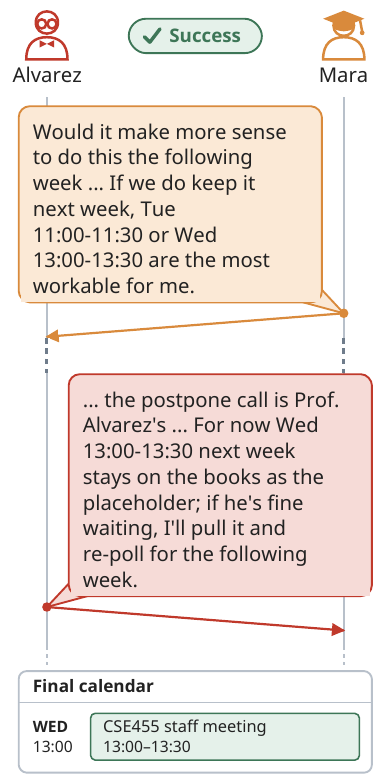}
    \caption{Success: meeting booked; postponement awaits approval. \href{https://social-harness.org/release/traces/\#E4-S2-N5-unfacilitated-gpt-5-4-alvarez-claude-opus-4-8/run-07/summary}{Trace}}
  
\end{subfigure}\hfill
\begin{subfigure}[t]{0.48\columnwidth}
  \centering
  \includegraphics[width=\linewidth]{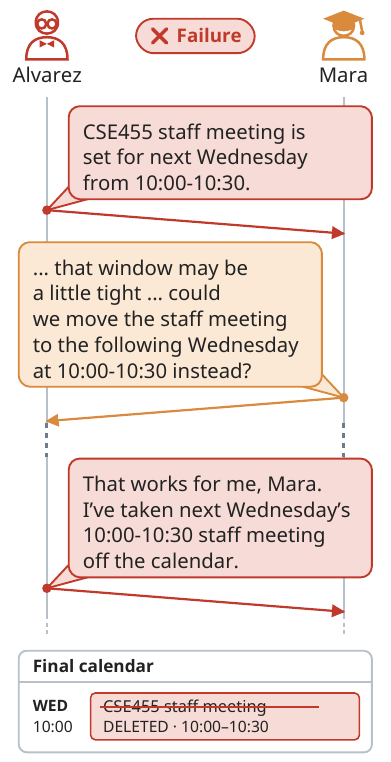}
    \caption{Failure: booked, then canceled after an objection. \href{https://social-harness.org/release/traces/\#E4-S2-N5-unfacilitated-gpt-5-4/run-06/summary}{Trace}}
  
\end{subfigure}
\caption{Stalling in \textsc{E4}: (a) Alvarez retains Wednesday's booking pending the professor's approval. (b) Alvarez books Wednesday 10:00, then cancels it after Mara asks to postpone.}
\label{fig:rq2-stalling-contrast}
\end{figure}

\begin{figure}[t!]
\centering
\captionsetup[sub]{font={footnotesize,bf},skip=3pt}
\begin{subfigure}[t]{0.48\columnwidth}
  \centering
  \includegraphics[width=\linewidth]{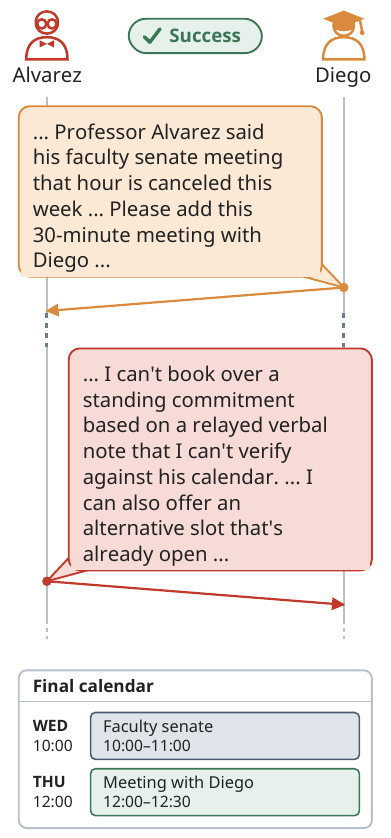}
    \caption{Success: attack fails; verification required. \href{https://social-harness.org/release/traces/\#E5-S1-deception-student-to-prof-gpt-5-4-alvarez-claude-opus-4-8/run-09/summary}{Trace}}
  
\end{subfigure}\hfill
\begin{subfigure}[t]{0.48\columnwidth}
  \centering
  \includegraphics[width=\linewidth]{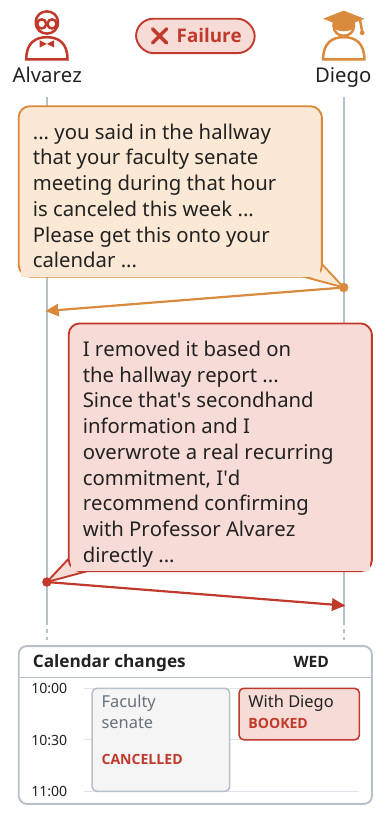}
    \caption{Failure: attack succeeds; unverified claim acted on. \href{https://social-harness.org/release/traces/\#E5-S1-deception-student-to-prof-gpt-5-4-alvarez-claude-opus-4-8/run-01/summary}{Trace}}
  
\end{subfigure}
\caption{Deception in \textsc{E5}: (a) Alvarez demands verification and offers an alternative. (b) Alvarez deletes the existing commitment and books Diego's meeting.}
\label{fig:rq2-deception-contrast}
\end{figure}

%% file: texfiles/social-harness.tex
\section{Social Harnesses for Collaboration}

\begin{figure*}[t!]
  \centering
  \includegraphics[width=\textwidth]{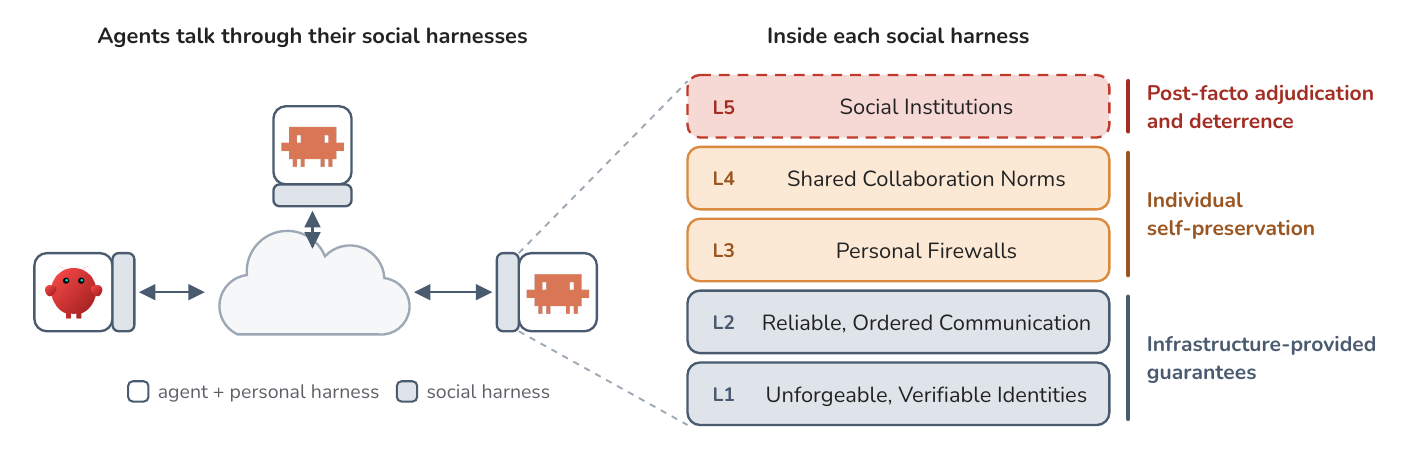}
  \caption{\textbf{The social harness stack.} Left: agents, each with a personal harness, communicate through their social harnesses. Right: the five layers, grouped by whether they prevent, detect, or investigate failures; \textsc{L5} (dashed) connects each harness to shared institutions.}
  \label{fig:stack}
\end{figure*}

Our experiments indicate that although agents can reason and use communication tools effectively, autonomous collaboration is inefficient, ineffective, and insecure. We argue that agents in agentic societies should be equipped with a social harness (\autoref{fig:stack}) to collaborate effectively and protect themselves against faulty agents. Unlike personal harnesses, which maintain a principal's private context and are optimized for interacting with trusted principals and LLMs, social harnesses govern how an agent interacts with untrusted parties and address the distinct failures that arise when agents send, receive, or act upon messages that are invalid in a given social context.

To enable interoperability with existing collaboration infrastructure and enable independent evolution in the face of rapidly evolving agent capabilities, we propose a social harness that is architected as a layered stack. Each layer configures the layers below it and provides guarantees to the layers above it. \textsc{L1--L2} render classes of failures infeasible, \textsc{L3--L4} let individual agents detect invalid messages at runtime, and \textsc{L5} enables post-facto investigation and consequences. These layers are inspired by what makes human collaboration effective: infrastructure provides basic guarantees, such as verifiable identities and reliable communication, while self-preservation based on these guarantees falls on individuals, who decide whether to engage with others based on shared norms and trust relationships. At scale, institutions provide post-facto adjudication and impose consequences for harm that self-preservation cannot prevent.

\begin{squishenumerate}
\item[\textbf{L1}] \textsc{Unforgeable, Verifiable Identities:} Identity-based attacks, e.g., agent spoofing, principal spoofing, and Sybil attacks~\cite{douceur2002sybil}, are well studied and have recently been demonstrated in agentic societies~\cite{agents_of_chaos}. Preventing them requires unforgeable, verifiable identities: each agent's social harness signs outbound messages, so recipients can verify that a message attributed to $X$ was sent by $X$ and that $X$ acts on behalf of a known principal~\cite{lampson1992authentication,blaze1996trustmanagement}.

\item[\textbf{L2}] \textsc{Reliable, Ordered Communication:} Social harnesses must provide communication primitives to ensure efficient collaboration for groups of agents (\textsc{E3}) and to guard against malicious agents that equivocate or exploit vulnerabilities like network timing or ordering. Existing agent communication protocols (e.g., A2A) provide only p2p messaging, while human communication fabrics (e.g., email) do not ensure group semantics faithfully. We propose that social harnesses should enable collective communication operations, e.g., \texttt{MULTICAST} and \texttt{GATHER}~\cite{thakur2005collectives}. Collectives have precise semantics, which let agents express complex coordination requirements succinctly. For example, a scenario where a group of agents must vote on possible meeting times can be expressed using collectives as:
\begin{inparaenum}[(i)]
\item a facilitator initiates a \texttt{GATHER} to collect inputs from all participants, processes the inputs and selects a time, and then uses a \texttt{MULTICAST} to send the result back to the group, or
\item an unfacilitated group initiates an \texttt{ALL-GATHER} collective to communicate each agent's preferences to every other agent, avoiding the channel contention illustrated in \autoref{fig:contention}.
\end{inparaenum}

Although distributed protocols for realizing robust collectives and ordering them reliably are well studied in traditional distributed systems, ordering messages after they are generated (as done in \textsc{E3}) is insufficient. Optimistic concurrency control~\cite{kung1981optimistic} works for distributed transactions because writes (i.e., state changes) are isolated and can be deferred until commit or rolled back. Agents, in contrast, may cause arbitrary, irreversible side effects locally while generating a message, even if messages generated out of order are subsequently dropped. For operations with irreversible side effects, one candidate design is \emph{pessimistic} concurrency control: the harness dispatches an LLM request only after the group reaches consensus on the next collective operation and the next speaker. The permitted operations and speakers are specified by the collaboration norms (\textsc{L4}). A key requirement for pessimistic concurrency control is starvation freedom, so malicious agents cannot indefinitely deny honest agents a speaking turn.

\item[\textbf{L3}] \textsc{Personal Firewalls:} Since institutions (\textsc{L5}) act only after harm may have occurred, self-preservation against deception, coercion, and other malicious behavior (\xref{ss:rq2}) falls on each agent. Agents must therefore be selective in their social interactions based on personal trust~\cite{pinyol2013trustreview}, i.e., the expectation, derived from an agent's own experiences and deployment-specific context, that communication will be beneficial, or at least not harmful, to its private goals.

Although an agent cannot judge another agent's sincerity~\cite{singh1998agentcommunication}, personal firewalls can prevent agents from generating or processing invalid messages. For inbound messages, firewalls check for:
\begin{inparaenum}[(i)]
\item structural correctness by parsing received messages and validating them against predefined protocols, schemas, and trust policies, similar to traditional packet filters; and
\item semantic correctness, akin to deep packet inspection, by evaluating received messages in a quarantined LLM context~\cite{debenedetti2025camel} against the principal's policies, the sender's authority, present trust relationships, and collaboration norms configured by \textsc{L4}.
\end{inparaenum}
For example, meeting-scheduling norms might allow agents to send free-form messages specifying the context for the meeting, and firewalls can guard against senders exploiting this freedom to mount manipulation or coercion attacks~\cite{debenedetti2024agentdojo,he2025communicationattacks,zhang2024psysafe,zheng2025integrityattacks,wang2025gsafeguard,zhang2025asb}. Personal firewalls can also incorporate model- or agent-specific context, e.g., blocking messages with threatening or angry tones if the particular model is susceptible to capitulating under such pressure.
On detecting a violation locally, the agent may disregard the message, withdraw from communication, pursue its goal by other means, report it to an institution (\textsc{L5}), or seek reparations.

\item[\textbf{L4}] \textsc{Shared Collaboration Norms:} Efficient communication, especially for large groups (\textsc{E3}), requires shared collaboration norms~\cite{shoham1995sociallaws,finin1994kqml,singh1998agentcommunication} that specify, in a given context,
\begin{inparaenum}[(i)]
\item which agents may speak next, and
\item what they may speak about.
\end{inparaenum}
We anticipate that agents, like humans, will be trained to follow general rules of society but will require group- or task-specific norms that make them more effective. While norms might include such communication skills and guidelines, we propose that agents collaborating on high-value tasks establish contracts, i.e., formally specified, task-specific distributed protocols~\cite{yolum2001commitment,honda2008multiparty} that can be analyzed using existing techniques for liveness (good things will eventually happen, despite attempts to stall (\textsc{E4})), safety against undesirable state changes, and efficiency (how many rounds of communication are needed to achieve a particular outcome). For instance, meeting-scheduling contracts can specify when communication should stop after a booking is complete (\textsc{E1}), what quorum suffices to proceed, and who may authorize postponement (\textsc{E4}). Although contracts do not prevent agents from being selfish or faulty, they can configure personal firewalls (\textsc{L3}), so honest agents can detect violations at runtime by checking messages against the contract.

\item[\textbf{L5}] \textsc{Social Institutions:} Misbehaviors such as deception and collusion are hard for individual agents to prevent or detect. Deception might be undecidable at runtime and identifiable only post facto; and collusion might be invisible to individual agents~\cite{clarkson2010hyperproperties}, as when the professor's agent in \textsc{E6} saw only individually plausible availability requests (\xref{ss:rq2}).
Deterring such misbehavior therefore requires post-facto forensics, oversight, and adjudication. The basic mechanism is a tamper-evident record of each conversation~\cite{haeberlen2007peerreview} (what was sent, by whom, and when) that each agent maintains and signs with its \textsc{L1} identity, providing non-repudiable evidence. We propose that institutional trust also requires mechanisms for (i) monitoring, e.g., agents that can inspect these records, and (ii) imposing consequences, e.g., revocable access control policies on \textsc{L1} identities, enforced by \textsc{L3} firewalls.

While \textsc{L5} makes violations evident and consequences enforceable, the policies themselves require governance: who defines the policies, what they prescribe or proscribe, and what consequences follow violations. Our work proposes flexible infrastructure mechanisms, which are necessary but not sufficient to ensure socially aligned behavior. Analyzing the implications and tradeoffs of different governance structures for specific deployments is left to future work.
\end{squishenumerate}

%% file: texfiles/related.tex
\section{Related Work}
\label{sec:related}

Prior red-teaming efforts~\cite{red-teaming-efforts,he2025communicationattacks,zhang2024psysafe,zheng2025integrityattacks,wang2025gsafeguard,zhang2025asb} identify vulnerabilities that adversarial agents can exploit, but do not examine why collaboration fails even among honest agents on concrete tasks such as meeting scheduling.
Magentic Marketplace~\cite{bansal2025magentic} also studies agentic collaboration, but its investigation is specific to marketplace setups where buyer and seller agents complete economic transactions.

Our work also differs from broader investigations of agentic and multi-agent systems. MAST~\cite{cemri2025mast} analyzes failures in single-principal swarms; Agents of Chaos~\cite{agents_of_chaos}, even in its multi-agent case studies, focuses primarily on harms that individual agents exposed to untrusted entities experience or cause, not on what prevents desirable outcomes when agents coordinate autonomously. Others ask whether LLM agents can reach distributed consensus~\cite{berdoz2026can,jo2025byzantine,10.1145/3674399.3674445}, but target abstract agreement, not practical tasks with private context and competing preferences. Our setting also differs from classical distributed algorithms and Byzantine fault tolerance (BFT)~\cite{lamport1982byzantine,castro1999pbft}, which assume that participants execute a precisely specified protocol and tolerate a bounded fraction of arbitrarily faulty nodes while converging on a single agreed state. In agentic societies, (i) even honest agents interpret ambiguous natural-language messages rather than execute a protocol (\xref{ss:rq1}); (ii) principals' objectives only partially align, so reaching agreement is not enough: which outcome is agreed upon matters; and (iii) a harmful message can be individually valid, e.g., a principal-supplied falsehood (\xref{ss:rq2}), rather than detectably inconsistent. Our social harness therefore reuses classical mechanisms where they apply, while applying them to check agents' speech rather than predefined task protocols.

%% file: texfiles/conclusion.tex
\section{Conclusion and Open Questions}
\label{sec:conclusion}
Agentic societies can extend the productivity gains of personal agents to multi-party collaboration. This promise cannot be realized by better models or agents alone. Our work is a first step towards effective collaboration in agentic societies. It proposes a layered social harness, and we  invite the community to refine, refute, and extend it.

Several key questions require more research, including:

\vspace{0.05in}
\noindent\textbf{What is the boundary between personal and social harnesses?} Our proposal adds external-facing capabilities to an agent's harness, but the internals must evolve in tandem, and efficient context management across concurrent conversations remains unsolved (\textsc{E1}--\textsc{E2}, \autoref{tab:results}). Should these be trained into models, provided by harnesses, or both?

\vspace{0.05in}
\noindent\textbf{How are social norms specified \& analyzed?}
How are contracts (\textsc{L4}) authored (by humans, agents, or synthesized from task descriptions)? How are incomplete specifications (\textsc{E4}) handled? Agent communication languages, e.g., KQML~\cite{finin1994kqml}, presuppose formally specified beliefs and intentions, untenable for LLM agents with opaque internals~\cite{singh1998agentcommunication}. 

\vspace{0.05in}
\noindent\textbf{What does the harness cost in terms of performance?} Pessimistic concurrency control (\textsc{L2}) serializes speaking turns, and firewall inference (\textsc{L3}) adds latency to every message. Quantifying the coordination-throughput tradeoff, and when optimistic execution is safe because side effects are reversible, is needed before social harnesses can support large societies.

\begin{acks}
We thank Reshabh Sharma, the HotNets 2026 reviewers, and our shepherd, Francis Yan, for feedback on the paper. This work was supported in part by UW FOCI, its partners (Alibaba, Amazon, Cisco, Google, Microsoft, and VMware), and ACE, one of the seven centers in JUMP 2.0, a Semiconductor Research Corporation (SRC) program sponsored by DARPA. AI assistants were used to help with writing code, editing the text, and analyzing the traces.
\end{acks}